\documentstyle[floats,prl,aps]{revtex} 
 
\input epsf

\begin{document}

\def\degrees{\hbox{${}^\circ$\hskip-3pt .}}
\def\spose#1{\hbox to 0pt{#1\hss}}
\def\simlt{\mathrel{\spose{\lower 3pt\hbox{$\mathchar"218$}}
        \raise 2.0pt\hbox{$\mathchar"13C$}}}
\def\simgt{\mathrel{\spose{\lower 3pt\hbox{$\mathchar"218$}}
     \raise 2.0pt\hbox{$\mathchar"13E$}}}

\newcommand{\Phigb}{\Phi_{\gamma b}}
\newcommand{\RF}{{\cal{T}}}
\newcommand{\Aa}{{\cal A}_a}
\newcommand{\Ab}{{\cal A}_b}
\newcommand{\bg}{{b\gamma}}
\newcommand{\eal}{\!\!\! & = & \!\!\!}

\title{TESTING INFLATION WITH SMALL SCALE CMB ANISOTROPIES$^{\rm\dag}$}
\author{Martin White${}^1$ \& Wayne Hu${}^2$}
\address{$^{1}$Enrico Fermi Institute, 5640 S.~Ellis Ave, Chicago,
IL 60637\\
$^{2}$Institute for Advanced Study, Princeton, NJ 08540}

\twocolumn[
\maketitle
\widetext

\begin{abstract}{\baselineskip 0.4cm
We discuss ability of the harmonic pattern of peaks
in the CMB angular power spectrum to test inflation.
By studying robust features of alternate models, which must all 
be isocurvature in nature, we reveal signatures unique to inflation.
Inflation thus could be validated by the next generation of 
experiments.}
\end{abstract}
\vskip 1.0truecm
]
\narrowtext


Inflation is the front running candidate for generating fluctuations in the
early universe: the density perturbations which are the precursors of galaxies
and cosmic microwave background (CMB) anisotropies today.
By ``inflation'' here we simply mean the idea that the
universe underwent a period of vacuum driven superluminal expansion during
its early evolution, which provides a mechanism of connecting, at early times,
parts of the universe which are currently space-like separated.
It has been argued that inflation is the unique {\it causal} mechanism for
generating correlated curvature perturbations on scales larger than the
horizon \cite{acausal,BigPaper}.
If there are unique consequences of such super-horizon curvature perturbations,
their observation would provide strong evidence for inflation.

\begin{figure*}[t]
\begin{center}
\leavevmode
\epsfxsize=3.25in \epsfbox{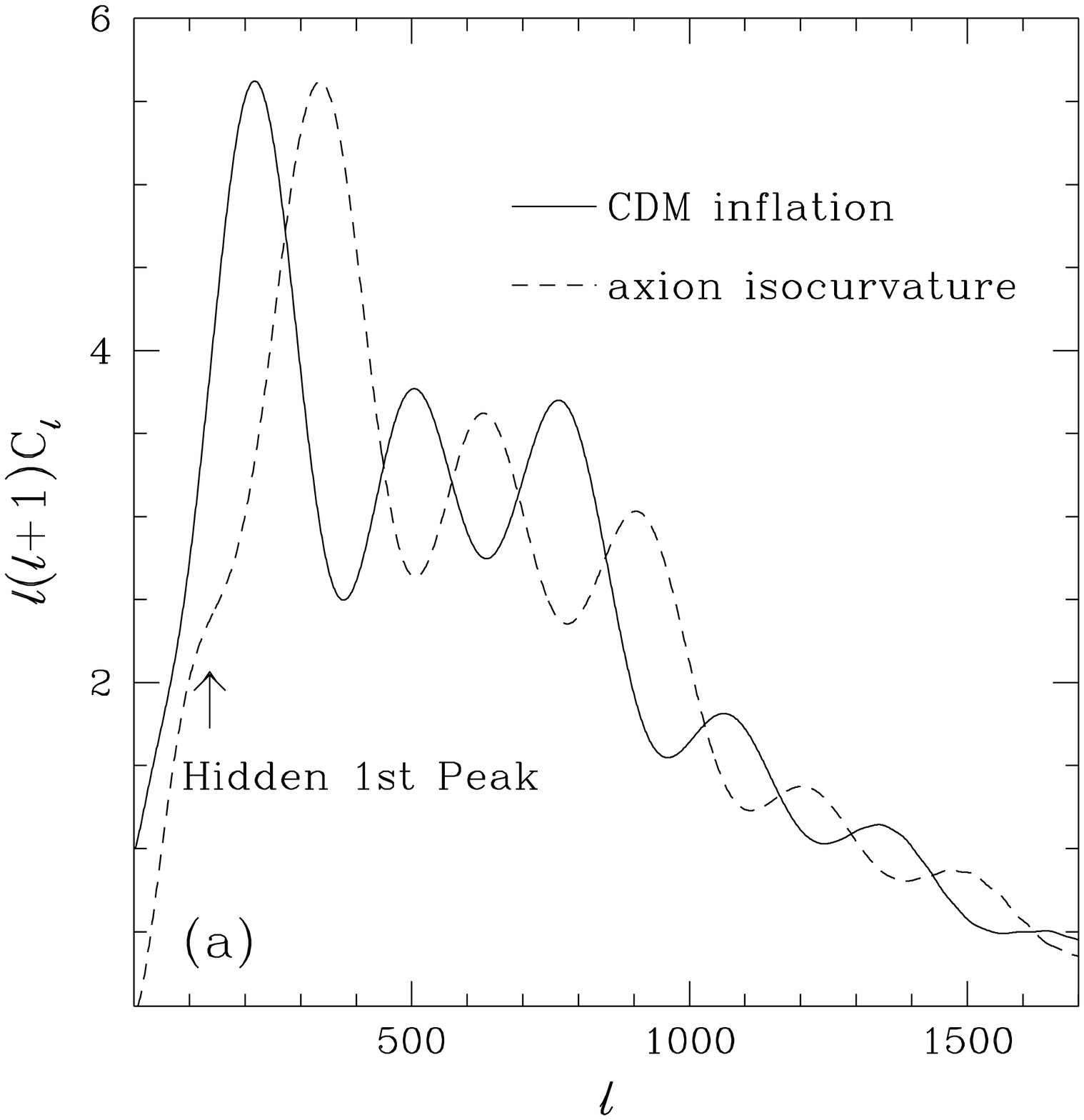}
\epsfxsize=3.25in \epsfbox{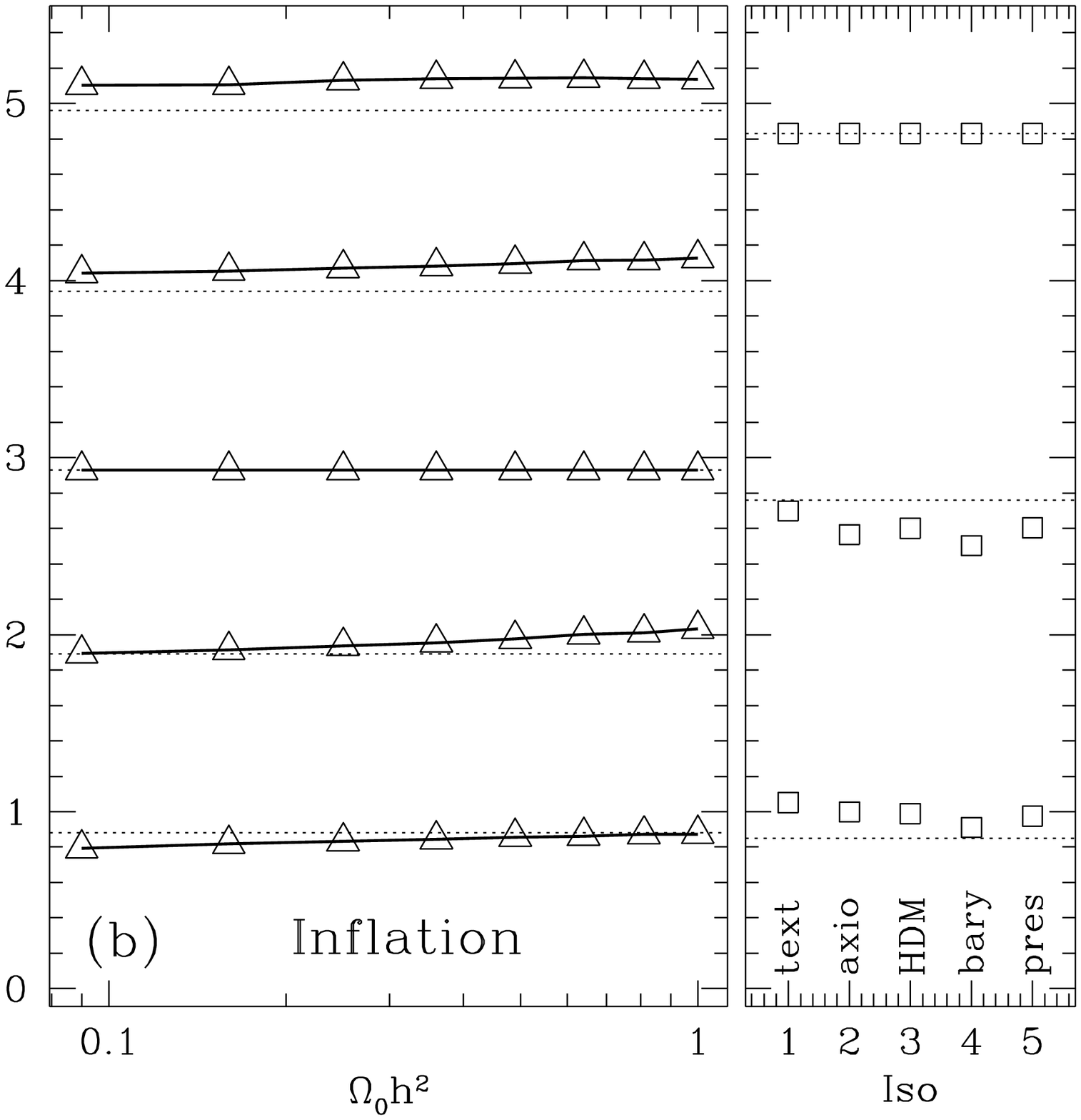}
\end{center}
\caption{\noindent (a) The angular power spectrum of a ``standard'' inflationary
CDM model with $\Omega_0=1$, $h=0.5$ and $\Omega_b=0.05$ (solid)
compared with an axionic isocurvature model (dashed) 
of the same parameters.
Note the peaks are offset from the inflationary prediction, and the first
``peak'' is more of a shoulder in this model.
(b) The {\it relative} positions of the peaks in the angular power spectrum
$\ell_1 : \ell_2 : \ell_3 \cdots $ for the inflationary 
(left panel, points) and 5 isocurvature models (right panel, points,
see text).
The series are normalized at $\ell_3$ to the idealized inflationary
and isocurvature series respectively (dotted lines). 
Test cases illustrate that the two cases remain
quite distinct, especially in the ratio of the first to the third peak and
to the peak spacing.}
\label{fig:scdm}
\end{figure*}

Here we probe the nature of the fluctuations through CMB
anisotropy observations of the acoustic signatures in the spectrum.
Many of the relevant technical details as well as more subtle examples can
be found in \cite{BigPaper}.
As a working hypothesis, we shall assume that the CMB spectrum exhibits a
significant harmonic signature: a series of peaks in the power spectrum when
plotted against multipole number $\ell$ (see Fig.~\ref{fig:scdm}a; for reviews
of the underlying physics of these peaks see \cite{HuSug,acoustic}).
Such a signature is expected in inflationary models and is characterized by
the locations and relative heights of the peaks as well as the position of
the damping tail.

The possibility of distinguishing some specific defect models from inflation
based on the structure of the power spectrum below $0\degrees5$ has recently
been emphasized \cite{CriTur,Albetal}.
By characterizing the features of such alternate models and revealing
signatures unique to inflation \cite{BigPaper}, we provide the
extra ingredients necessary to allow a test of the inflationary paradigm.
Another means of testing inflation is the consistency relation between the
ratio of tensor and scalar modes and the tensor spectral index
\cite{Davetal,KnoTur}.  However this test requires a large tensor signal
\cite{KnoTur} or it will be lost in the cosmic variance.
\footnotetext{$^{\rm\dag}$To appear in Proceedings of the 
XXXIth Moriond Meeting, {\it Microwave Background Anisotropies}. 
IASSNS-AST-96/37}

In Fig.~\ref{fig:scdm}a (solid lines), we show the angular power spectrum of
CMB anisotropies for a standard cold dark matter (CDM) inflationary model,
as a function of multipole number $\ell\sim\theta^{-1}$.
Let us review the physics behind the features in the spectrum below
$0\degrees5$ ($\ell\simgt200$).
Consider the universe just before it cooled enough to allow
protons to capture electrons.
At these early times, the photons and baryon-electron plasma are tightly
coupled by Compton scattering and electromagnetic interactions.
These components thus behaved as a single `photon-baryon fluid' with the
photons providing the pressure and the baryons providing inertia.
In the presence of a gravitational potential, forced acoustic oscillations in
the photon-baryon fluid arise.
The energy density, or brightness, fluctuations in the photons are seen by the
observers as temperature anisotropies on the CMB sky.
Specifically, if $\Theta_0$ is the temperature fluctuation $\Delta T/T$ in 
normal mode $k$, the oscillator equation is
\begin{equation}
{d \over d\eta}\left[ m_{\rm eff}{d\Theta_0\over d\eta}\right] +
  {k^2 \over 3}\Theta_0
  = -F[\Phi,\Psi,R]
\label{eqn:Oscillator}
\end{equation}
with
\begin{equation}
F[\Phi,\Psi,R]= {k^2 \over 3}m_{\rm eff}\Psi +
  {d \over d\eta}\left[m_{\rm eff}{d\Phi\over d\eta}\right],
\end{equation}
where $m_{\rm eff}=1+R$, $R=3\rho_b/4\rho_\gamma$ is the baryon-to-photon
momentum density ratio, $\eta=\int dt/a$ is conformal time,
$\Phi$ is the Newtonian curvature perturbation, and 
$\Psi \approx -\Phi$ is the gravitational potential \cite{BigPaper,HuSug}.

\begin{figure}
\begin{center}
\leavevmode
\epsfxsize=3.25in \epsfbox{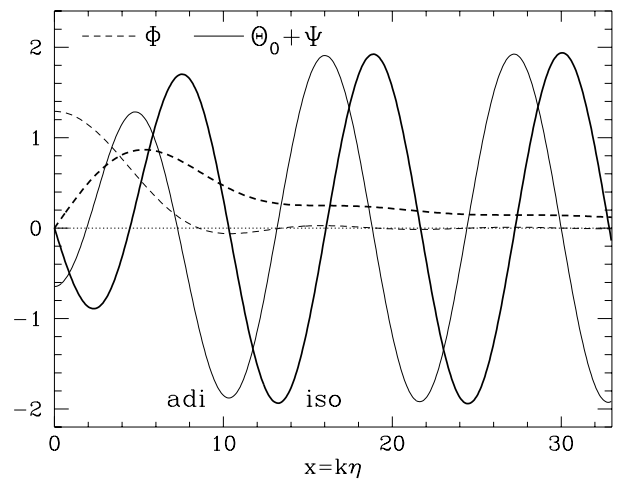}
\end{center}
\caption{\noindent The self-gravity of the photon-baryon fluid drives a
cosine oscillation for adiabatic initial conditions (thin lines) and a
sine oscillation (thick lines) for isocurvature initial conditions.
The dashed lines show the full potential, the solid lines the effective
temperature.}
\label{fig:drive}
\end{figure}

In an inflationary model, the curvature or potential fluctuations are created
at very early times and remain constant until the fluctuation crosses the
sound horizon.  Inside the sound horizon the pressure becomes important and
the potential begins to decay (see Fig.~\ref{fig:drive}).
As a function of time, this force excites a cosine\footnote{The inflationary
series only reaches a cosine asymptotically at high peak number:
$0.88:1.89:2.93:\cdots$.  Likewise the most natural isocurvature series starts
at $0.85:2.76:4.83:\cdots$ \cite{BigPaper}.}
mode of the acoustic oscillation with peaks at
$x/\sqrt{3}\simeq\pi, 2\pi, 3\pi, \cdots$.
The first feature represents a compression of the fluid inside the potential
well as will become important in the discussion below.
Furthermore, the harmonic series of acoustic peak location
$\ell_1:\ell_2:\ell_3 $ approximately follows the cosine series of extrema
$1:2:3\cdots$.
There are two concerns that need to be addressed for this potential test of
inflation.  How robust is the harmonic prediction in the general class of
inflationary models? Can any other model mimic the inflationary series?  

The peak ratios are not affected by the presence of spatial curvature or
a cosmological constant in the universe \cite{BigPaper}.  However it is
possible to distort the shape of the first peak by non-trivial evolution of
the metric fluctuations after last scattering.   
For example, the magnitude of the scalar effect increases with the influence of
the radiation on the gravitational potentials, e.g.~by a decrease in the matter
content $\Omega_0h^2$.
Tensor fluctuations could distort the first peak and spectral tilt shift the
series only if they are very large.  That possibility is inconsistent with the
observed power at degree scales.\footnote{If degree scale power is present,
$n\ge0.5$, the peaks and positions are not obscured by tilt.}
The damping of power in the oscillations at small scales due to photon
diffusion cuts off the spectrum of peaks and could also confuse a measurement
of their location.
In Fig.~\ref{fig:scdm}b, we plot the ratio of peak locations as a function of
$\Omega_0 h^2$.
Although the first peak is indeed slightly low in the low $\Omega_0h^2$ models
the harmonic series is still clearly discernible in the regular spacing of the
higher peaks.  Two numbers serve to quantify the spectrum: the
ratio of third to first peak location $\ell_3/\ell_1 \approx 3.3-3.7$
and the first peak location to the spacing between the peaks 
$\ell_1/\Delta \ell \approx 0.7-0.9$.  Ratios in this range are 
a robust prediction of inflation with reasonable baryon
content.\footnote{If {\it both} the baryon content $\Omega_bh^2\simgt0.03$ and
the CDM content $\Omega_0 h^2\simgt0.6$ are anomalously
high then the second peak (r) will be hidden by the baryon inertia
\cite{BigPaper}.}


Is the cosine harmonic series a {\it unique} prediction of inflation?
Causality requires that all other models form significant curvature
perturbations near or after horizon crossing.\footnote{This does not
preclude the possibility of white noise curvature perturbations at low
$k$ generated by pressure perturbations.  See \cite{BigPaper,HuSpeWhi} for
more discussion.}
We call these {\it isocurvature} models.  The axionic isocurvature model
of Fig.~\ref{fig:scdm}a 
(dashed lines) is representative \cite{KawSugYan}.  
Since curvature fluctuations start out small and grow until horizon crossing,
the peak locations are phase shifted with respect to the inflationary
prediction (see Fig.~\ref{fig:drive}).

In typical models, including 
the baryon isocurvature \cite{HuSug},
texture \cite{CriTur}, 
axionic isocurvature \cite{KawSugYan}, 
hot dark matter isocurvature \cite{deLSch}, and pressure scaling
\cite{HuSpeWhi} models,
the peaks approximately form a sine series $1:3:5\cdots$
(see Fig.~\ref{fig:scdm}b).
The origin of this sine series signal is described in \cite{BigPaper} where
it is shown acoustic oscillations are driven as the fluctuation passes the
sound horizon by the fluid's own self gravity.
The first ``peak'' is a rarefaction, which is a continuation of the
super-horzon scale behaviour of the isocurvature model.  This 
is robust due to causal constraints as we shall see below and also
implies that the first peak can be quite small.  
Thereafter, the photon pressure becomes important
and the fluid collapses into the potential, becoming more dense and boosting
$\Phi$ (see Fig.~\ref{fig:drive}).  This enhances the second peak.
The tendency to first fight and then help the driving potential is generic.
Because this is a resonant process, in most cases it dominates over other
truly external effects.  In particular, all models in which fluctuations are
generated causally by pressure perturbations that are constrained to produce
nearly scale invariant CMB anisotropies will be dominated by this effect in
their acoustic signature \cite{HuSpeWhi}.  

However since we wish to test inflation against all possible alternatives,
let us now turn to the broader class of isocurvature models.
Isocurvature models with more radical source behavior may introduce some 
other phase shift with respect to
the inflationary prediction. Might this allow an isocurvature scenario 
to mimic the inflationary prediction?
Two possibilities arise.  If the first isocurvature feature, which is
intrinsically low in amplitude, 
is hidden, e.g.~by external metric fluctuations
such as tensor and vector contributions between last scattering and the
present, the series becomes approximately $3:5:7$.
Might this be mistaken for an inflationary spectrum, shifted to smaller angles
by the curvature of the universe?
The spectra remain distinct since the spacing between 
the peaks $\Delta\ell$ is 
model-independent: it reflects the natural period of the oscillator.
The ratio of the first peak position to peak spacing $\ell_1/\Delta\ell$ is
thus larger by a factor of $1.5$ in this case if $\Omega_0 h^2$ is fixed.
In \cite{BigPaper}, we treat the ambiguity that arises if this and other
background quantities are unknown.
More generally, any isocurvature model that either introduces a pure phase
shift or generates acoustic oscillations only well inside the causal horizon
can be distinguished by this test.  Of course, isocurvature models need not
exhibit a simple regularly-spaced series of peaks \cite{BigPaper,Albetal},
but these alternatives could not mimic inflation.



The remaining possibility is that an isocurvature model might be tuned so that
its phase shift precisely matches the inflationary prediction.
Heuristically, this moves the whole isocurvature spectrum in Fig.~1 toward
{\it smaller} angles.  We shall see that causality forbids us to make the
shift in the opposite direction.
As Fig.~\ref{fig:scdm}a 
implies, the relative peak heights can distinguish this possibility
from the inflationary case.

The important distinction comes from the process of compensation, required
by causality.
During the evolution of the universe, the dominant dynamical component
counters any change in the curvature introduced by an arbitrary source
\cite{BigPaper}.
Producing a positive curvature perturbation locally stretches space.  The
density of the dominant dynamical component is thus reduced in this region,
and hence its energy density is also reduced.  This energy density however
contributes to the curvature of space, thus this reduction serves to offset
the increased curvature from the source.
Heuristically, curvature perturbations form only through the motion of
matter (see \cite{HuSpeWhi} for more details), which causality forbids
above the horizon.

In the standard scenario, the universe is radiation dominated when the smallest
scales enter the horizon (see \cite{BigPaper} for exotic models).
Thus near or above the horizon, the photons resist any change in curvature
introduced by the source.  Breaking $\Phi$ into pieces generated by the
photon-baryon fluid ($\gamma b$) and an external source ($s$), we find
in this limit \cite{BigPaper}
\begin{equation}
x^2 \Phigb'' + 4 x \Phigb' = - x^2 \Phi_s'' - 4 x \Phi_s' ,
\label{eqn:Compensationa}
\end{equation}
where primes denote derivatives with respect to $x = k\eta$.

Thus the first peak in an isocurvature model, if it is sufficiently close to
the horizon to be confused with the inflationary prediction and follows the
cosine series defined by the higher peaks, must have photon-baryon fluctuations
anti-correlated with the source.  
The first peak in the rms temperature thus represents the rarefaction (r)
stage when the source is overdense rather than a compression (c) phase as
in the inflationary prediction.  The peaks in the inflationary
spectrum obey a c-r-c pattern while the isocurvature model displays a r-c-r
pattern.  Though compressions and rarefactions have the same amount of
power (squared fluctuation), an additional effect allows us to
distinguish the two: baryons provide extra inertia to the photons to which
they are tightly coupled by Compton scattering (the $m_{\rm eff}$ terms in
Eq.~\ref{eqn:Oscillator}).
If overdense regions represent gravitational wells,\footnote{Neil Turok
(private communication) has
shown that for a particular choice of stress-energy tensor, with an
anisotropic stress large compared with the density, it is possible to have
{\it under}densities associated with potential wells.  If such a model
also has peaks $\pi$ out of phase with the cosine mode it could mimic an
inflationary spectrum.} this inertia enhances compressions at the expense
of rarefactions leading to an alternating series of peaks in the rms
\cite{HuSug}.
For reasonable baryon content, the {\it even} peaks of an isocurvature model
are enhanced by the baryon content whereas the {\it odd} peaks are enhanced
under the inflationary paradigm (see Fig.~\ref{fig:scdm}a).
This non-monotonic modulation of the peaks is not likely to occur in
the initial spectrum of fluctuations.
The oscillations could be driven at exactly the (evolving) natural frequency
of the oscillator in such a way as to counteract this shift, but such a long
duration tuned driving seems contrived.

There is one important point to bear in mind.  Since photon diffusion damps
power on small scales, the 2nd compression (3rd peak) in an inflationary model
may not be {\it higher} than the 1st rarefaction (2nd peak), even though it is
enhanced (see e.g.~Fig.~\ref{fig:scdm}a).
However it will still be anomalously high compared to a rarefaction peak,
which would be both suppressed by the baryons and damped by diffusion.
Since the damping is well understood this poses no problem in principle
\cite{BigPaper}.

Diffusion damping also supplies an important consistency test.
The physical scale depends only on the background cosmology and not on the
model for structure formation (see Fig.~\ref{fig:scdm}a 
and \cite{HuSug,BigPaper}).
This fixed scale provides another measure of the phase shift introduced by
isocurvature models.  For example, if the first isocurvature peak in 
Fig.~\ref{fig:scdm}a 
is hidden, the ratio of peak to damping scale increases by a factor of 1.5
over the inflationary models.  We also consider in \cite{BigPaper} how the
damping scale may be used to test against exotic background parameters and
thermal histories.


In summary, the ratio of peak locations is a robust prediction of inflation.
If acoustic oscillations are observed in the CMB, and the ratio of the 3rd to
1st peak is not in the range $3.3-3.7$ or the 1st peak to peak-spacing in
the range $0.7-0.9$ then either inflation does not provide the main source of
perturbations in the early universe or big bang nucleosynthesis grossly
misestimates the baryon fraction.
The ranges can be tightened if $\Omega_0 h^2$ is known.
If the spatial curvature of the universe vanishes, these tests require CMB
measurements from $10-30$ arcminutes.  Even if the location of the first peak
is ambiguous, as might be the case in some isocurvature models, these tests
distinguish them from inflation.
Isocurvature models thus require fine tuning to reproduce this spectrum.
To close this loophole, the relative peak heights can be observed.  If
the location of the peaks follows the inflationary prediction, the enhancement
of odd peaks is a unique prediction of inflation.\footnote{This is strictly
only true if the photon energy density is significant when the relevant scales
enter the horizon and gravitational potential wells are associated with
{\it over}dense regions.}


The true discriminatory power of the CMB manifests itself in the spectrum as
a whole, from degree scales into the damping region.
In particular, we emphasize the acoustic {\it pattern} which arises from
forced oscillations of the photon-baryon fluid before recombination, including 
the model-independent nature of the damping tail.
The tests we describe rely on the gross features of the angular power spectrum
and so could be performed with the upcoming generation of array receivers and
interferometers.

\acknowledgements{We acknowledge useful conversations with J. Bahcall,
P.~Ferreira, A.~Kosowsky, J.~Magueijo, A.~Stebbins and M.~Turner.
We thank R.~Crittenden, A.~de Laix, and N. Sugiyama for supplying 
isocurvature power spectra. 
W.H.~was supported by the NSF and WM Keck Foundation. 
We would also like to thank F. Bouchet and all the organizers for
a pleasant and productive conference.}

\end{document}